# Structural Studies on Semiconducting Hydrogenated Amorphous Silicon Oxide Films

S. M. Iftiquar
Energy Research Unit
Indian Association for the Cultivation of Science
2 Raja S. C. Mullick Road, Jadavpur, Calcutta 700 032, INDIA

## Abstract

In hydrogenated amorphous silicon oxide ( a-SiO:H ) films, incorporation of oxygen enhances optical gap due to a large number of St-O-Si bond formation, which lies deep into valence band states. An induction effect of this Si-O on other bonds within the network also takes place. At higher oxygen content micro-void forms and bonded hydrogen accumulates in di and/or polyhydride form. At this stage a phase separation of Si-rich and O-rich region taking place. A peak shift of absorption spectra within 1850 - 2250 $cm^{-1}$ , towards higher wave number is continuous. A gradual increase and broadening of 850 cm-1 absorption band on both sides of peak position indicate higher structural disorder in network formation. It may be considered that the stretching vibration of-OH bonded to Si gives rise to 780 $cm^{-1}$ absorption band. This Si-OH formation is beneficial which prevents deterioration in photosensitivity due to reduction in bonded hydrogen content. Hydrogen content is found reducing as oxygen content increases from zero to ~15 at.%. A systematic study is carried out to correlate the optoelectronic property with local atomic arrangement.

# § 1. Introduction

Presence of O in a–Si:H network in a controlled manner raises its optical gap ( Eg ) as well as photo conductivity ( $\sigma_{ph}$ ), ( Hamakawa, Fujimoto, Okuda, Kashima, Nomura and Okamoto 1983 ), as compared to those for silicon carbide of similiar Eg. Carbon in hydrogenated amorphous silicon carbide ( a–SiC:H ) alloy semiconductor is intentionally added to serve the purpose of widening its band gap ( Tamada, Okamoto and Hamakawa (1981 ). But recently Fujikake, Ohta, Sichanugrist, Ohsawa, Ichikawa and Sanai 1994, has fabricated multijunction a–Si:H solar cells using boron–doped hydrogenated amorphous silicon oxide (a–SiO:H ) material as the window layer and intrinsic a–SiO:H as active material of the first cell and obtained 10.5% conversion efficiency. Before that Sindoh, Haga, Yamamoto, Murakami, Kumano and Watanabe 1991, successfully fabricated a–Si:H/a–SiO$_{(x)}$:H double layer photo diode as an image sensor.

Apart from technological interest, proper understanding of the material is necessary to improve knowledge about a–SiO:H prepared by ( RF PECVD ).

The X–ray photoelectron spectroscopic ( XPS ) analysis shows that the Si core level spectra contains five different energy levels symbolising different atomic environments of the Si atoms ( Watanabe, Haga and Lohner 1993 ). This X–ray photoelectra of the a–SiO:H, prepared from $CO_2$ and $SiH_4$ gas mixture in the rf glow discharge and low power regime of PECVD, revealed that it contains $Si–Si_{(4-n)}–O_n$; ( for n = 0 to 4 ), five components, that is five different possible combinations of Si and O in tetrahedral bond formation ( Watanabe et. al. ( 1993 ). Morimoto, Noriyama and Shimuzu 1987, prepared a–SiO:H by rf glow discharge of $SiH_4$ and $O_2$ and rf magnetron sputtering of Si and $SiO_2$ composites in the ambient of an Ar or Ar + $H_2$ gas mixture. They found that asymmetric stretching vibration of O in Si–O–Si is influenced by total oxygen content of the film. Depending upon atomic environment around any Si atom vibration of O will be modified and the resonance absorption band will get broadened due to such environmental disorder.

Smith and Angelotti 1959, observed that there is a peak shift associated to the Si–H vibrational frequency which is due to the change in interatomic spacings between Si and H in Si–H bond. When there is a Si–O bond formation, a charge transfer from Si to O between 0.51e to 0.65e occurs, Nucho and Madhukar 1980, ( where e is electronic charge ), due to higher Pauli electronegativity of O. This partial polarization of Si–O bond to $Si^{\delta+}$—$O^{\delta-}$, affects the $Si^{\delta+}$—H stretching frequency. From classical point of view, it is said that this $Si^{\delta+}$—H bonding force constant attains some higher reassigned value due to presence of O, ( Cardona 1983 ), which produces evolution of another absorption peak $\sim$2090 cm$^{-1}$ with increasing oxygen content, C( O ) of the film.

As oxygen is moderately added to a–Si it leads to formation of Si–O–Si structure within the network and an IR absorption band around 1020 cm$^{-1}$ appears due to asymmetric stretching vibration of O along the line joining the two Si–atoms to which it is bonded.

There appears other IR absorption peaks like that at 475 cm$^{-1}$. It is due to out of plane rocking vibration of O in Si–O–Si structure ( Lucovsky 1983 ). 780 cm$^{-1}$ peak is due to strongly coupled vibration of Si–H and Si–O–Si ( Lucovsky 1982 ). Si–H bond bending oscillation shifts from 790 cm$^{-1}$ to 875 cm$^{-1}$ with increase in oxygen content ( Tsu, Lucovsky and Davison 1989 ).

In this report we present different IR absorption bands obtained from Fourier Transform infrared ( FTIR ) spectrometry and compare these to resonance vibrational frequencies of different atoms calculated for differing atomic environments. Optical and electrical properties are observed to be modified due to structural arrangements. Attempt has been made to correlate the optical gap and electronic properties of a–SiO:H films with the structural modifications.

According to the two phase model, if a–SiO:H be a mixed phase material constituting Si–rich phase, having mostly Si–Si bond, and oxygen rich phase, composed of mostly Si–O–Si rich region, in small clusters, then Si–rich region helps towards higher photoconductivity ( $\sigma_{ph}$ ) and oxygen rich region contributes to the widening in optical gap ( Watanabe, et. al. 1993 ). We avoid explaining from IR spectra whether our material follows Mixture Model ( Hubner 1980 ), ( George and D'Antonio 1979 ) or Random Bonding Model ( Watanabe, et. al. 1993 ) of bond formation among the constituent atoms. We do not also attempt to explain from IR analysis whether the nature of IR absorption spectra implies anything like the material is a mixed phase one or it is phase separated one ( Temkin 1975 ).

In the `theoretical' section we try to explain the situations that arise due to structural changes or, more specifically, due to the presence of oxygen atom.

## § 2. Theoretical
## 2.1. Oxygen Bonding

In amorphous network any one Si atom may be bonded to other Si, O and/or H at any possible number to fulfil its four valency or may have unfulfilled bond( s ). If one Si be bonded to two O atoms in Si–O–Si–O–Si type structure then each O of Si–O–Si site will have a contribution of IR absorption within 900 – 1200 $cm^{-1}$ region due to its asymmetric stretching vibration. But the peak position will shift towards higher wave number due to inductive effect of one Si–O– unit on the central Si atom. Due to this the frequency of vibration of the two oxygen atoms will be higher. That is if the two Si atoms of Si–O–Si do not contain any other oxygen atom bonded to it the asymmetric stretching vibration frequency of O will lie around 980 $cm^{-1}$ ( Lucovsky, Yang, Chao, Tyler and Czubatyj 1983). Presence of other O atom puts some inductive effect to shift the frequency to a higher value ( Akiharu Morimoto, et. al. 1987 ).

Another reason of shifting the peak position is due to bond angle variation at `bridging' O site. If the ( dihedral ) bond angle ( $\theta$ ) between the bonds of oxygen be more then the frequency of vibration will be high and vice versa, it is partly explained in figure 6. 1. According to electron scattering studies on a–$SiO_2$ it is found that the bond angle variation is existent, where $120^0 \le \theta \le 180^0$ and most probable value of which is ~$144^0$ ( Mozzi and Warren 1969 ).

The replacement of a Si atom by O in a–Si:H network induces certain change in atomic arrangements compared to those of a–Si:H due to its different electronegativity, valency number and dihedral angle $\theta$. At $\theta = 120^0$ the asymmetric stretching vibration includes bond bending due to the motion of oxygen atom, but with $\theta = 180^0$ no such bond bending takes place due to stretching vibration. Frequency in the second case will be higher than that in the first one.

If each of the two Si atoms to which the O atom is bonded, contains three other Si neighbour and if ideally it is possible to prepare such a material having only this type of oxygen bonding environment then the IR absorption band will have a definite Gaussian shape which is symmetrical around its peak position. This is not the case in reality. Atoms in a–SiO:H can be arranged in a random fashion and Si–O–Si is one such possibility. It can have any possible near neighbour and the angle $\theta$ may have wide ranging values. These fluctuations cause asymmetric stretching vibration of O to result in a broad absorption band within 900 to 1200 $cm^{-1}$ wavenumber region. To separate out each Gaussian peak and study them individually the spectra needs to be deconvoluted for its constituent absorption bands.

Stretching vibration frequency is calculated by the method given by Cardona (Cardona 1983 ) and the force constant values obtained from standard references are utilized.

The Si–O force constant is $6.043 \times 10^5$ dyn/cm ( Venkateswarlu and Sundaram 1955 ). When H is bonded to an O forming Si–OH, then this OH may oscillate, corresponding to the Si–O bonding force, with a resonance frequency 776 $cm^{-1}$. As this –OH of Si–OH vibrates the mass of Si atom will not influence the reduced mass of oscillation because the Si atom is bonded to the network by other three of its tetrahedral bonds, it is rigid to the network and oscillation of this Si atom is negligible.

We call the broad absorption band within 900 to 1200 cm$^{-1}$ as STO as it arises due to asymmetric stretching vibration of O in Si–O–Si.

We also assign 860 cm$^{-1}$ frequency as BO because of in plane bending mode oscillation of O in Si–O–Si and perpendicular to the line joining the two Si atoms, the 780 cm$^{-1}$ peak as due to the stretching mode of oscillation of OH in the Si–OH configuration.

## § 3. Experimental

We started analysing the material after depositing the film in a RF PECVD system using SiH$_4$, CO$_2$ and H$_2$ source gases. Films were deposited on crystalline Si ( c–Si ) substrates. A 13.56 MHz rf power was applied on two capacitively coupled planar electrodes. Base vacuum of the order of 10$^{-6}$ Torr was maintained prior to each deposition to ensure good vacuum sealing and minimum possibility for out of control contamination on the films deposited.

The deposited samples are characterized for optical gap, electronic properties and the structural analysis is done with the help of Fourier transform infrared ( FTIR ) spectrophotometer.

The FTIR measurements are carried out in a Perkin Elmer FTIR spectrophotometer ( Model No.– 1700 ) having a resolution of 1cm$^{-1}$.

The thickness of the samples, d , on c–Si are measured by a stylus type instrument. The d is kept around 6000 Å by controlling deposition time for each sample. This sample thickness is used to evaluate the absorption coefficients ( β in cm$^{-1}$ unit ) by equation ( 3 ).

$$\beta = ( 2.303 / d )( abs ) \qquad\qquad (1)$$

where, abs = thickness dependent absorption values obtained from FTIR spectrophotometer

The c–Si on which the sample is deposited imposes extra IR absorption to that of the sample, which is eliminated by taking a full range spectra of the corresponding reference c–Si substrate and subtracting it from that of the sample.

Atomic percent ( at.% ) of oxygen ( incorporated into a–Si:H network forming Si–O–Si structure), C( O ), is obtained from the integrated absorption strength of the IR absorption in the range of ( 900–1200 ) cm$^{-1}$ by the formula ( 2 ) ( Lucovsky, Yang, Chao and Czubatyj 1983 ),

$$C( O ) = A( O ) \, I( 980 \text{ cm}^{-1} ) \qquad\qquad (2)$$

where A( O ) = 0.156 at.%/( eV.cm$^{-1}$ ).

I( 980 cm$^{-1}$ ) = integrated absorption ( area under the absorption curve in wave number axis, in units of eV cm$^{-1}$ ) within the 900 –1200 cm$^{-1}$ band.

Atomic percent of hydrogen C( H ), is determined using equation ( 3 )

$$C( H ) = A( H ) \, I( 2000 \text{ cm}^{-1} ) \qquad\qquad ( 3 )$$

where A( H ) = 0.77 at.%/( eV.cm$^{-1}$ )

I( 2000 cm$^{-1}$ ) = integrated absorption ( in units of eV.cm$^{-1}$ ) within 1850–2250 cm$^{-1}$ region.

Optical gap is determined by using films deposited on Corning 7059 glass and evaluated from optical density data, obtained from UV–Visible spectrophotometer. It is observed that as we go on increasing the CO$_2$ flow during the deposition, from sample to sample, the optical gap increases.

# § 4. Results

Thin film a–SiO:H is a highly photosensitive semiconductor for C( O ) < 25 at.%. As oxygen atoms are introduced into the a–Si it brings about one more inhomogeneity in network formation. Bond formation between Si , O is stronger than that between Si , Si because of their higher electronegativety difference. The influence of oxygen towards photoconductivity $\sigma_{ph}$ , $\sigma_0$ , Eg , etc. is complex and the effect on these observables are stated occassionally, to point out the importance of certain IR results.

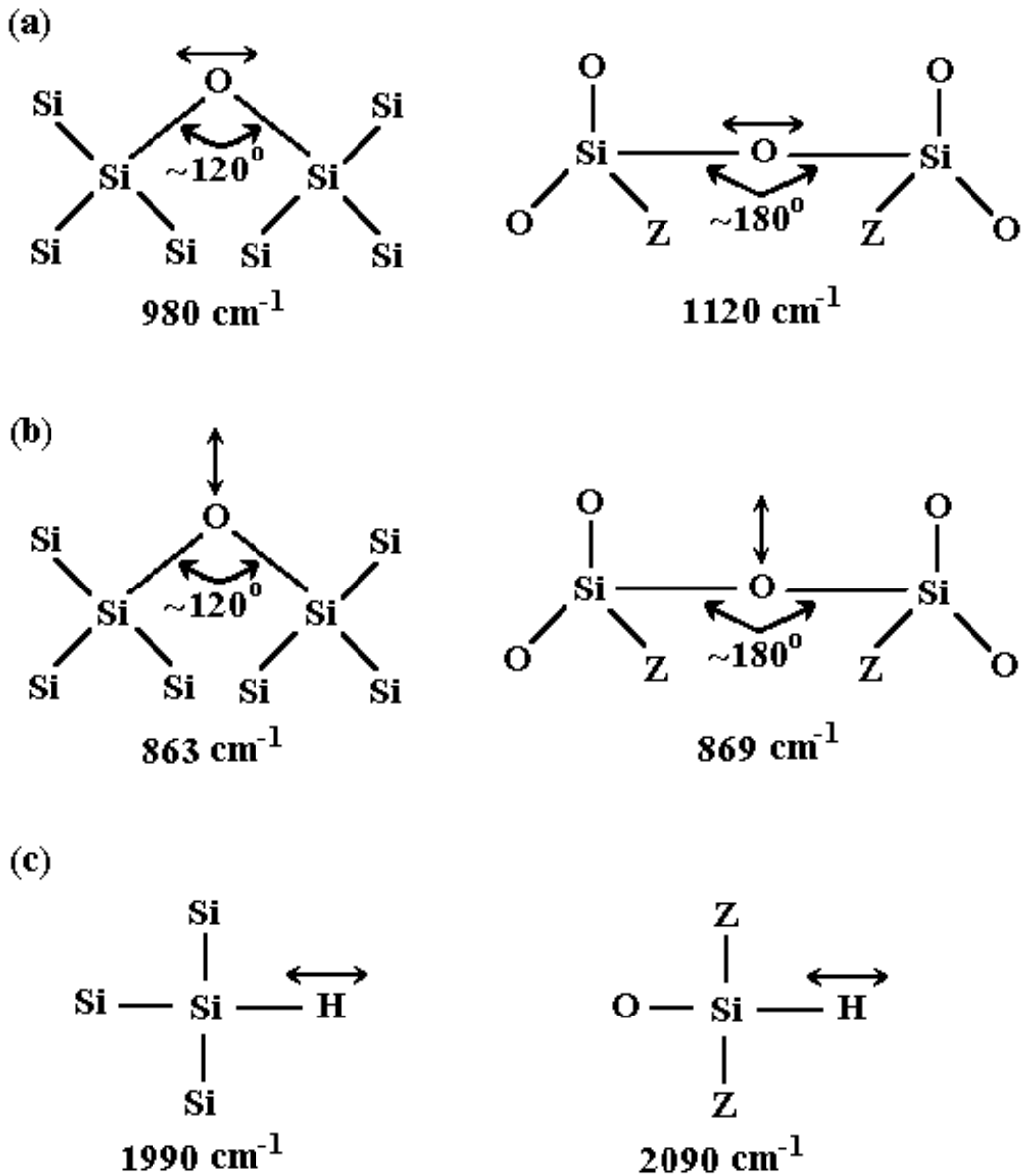

**(d)**

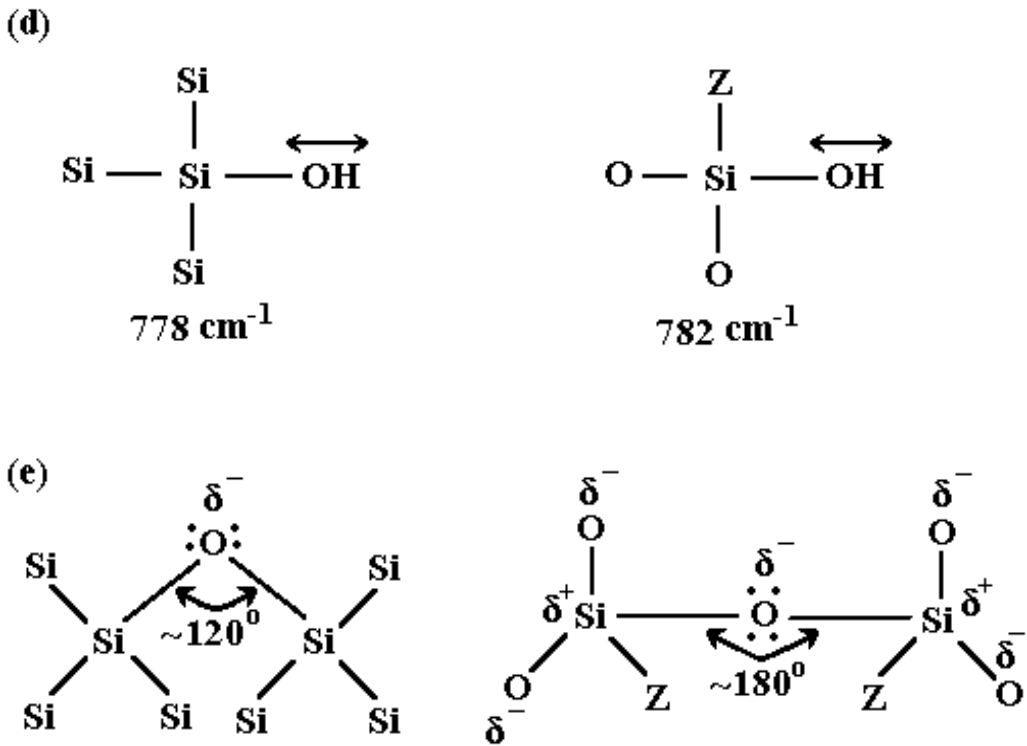

**Figure 1.** In this figure three dimentional structure is demonstrated in two dimension. Bond length does not have any significance but indicated bond angle are described. From figure 6. ( a ) to ( d ), schematically it is indicated the possible mode of vibration, that is detected in the IR study for different atoms or group of atoms, due to different bond angle θ or local atomic environment. Z stands for either Si or O atom. In ( a ) the inductive effect on asymmetric stretching vibration frequency of O is indicated. In ( b ) the amplitude of vibration of oxygen in bending mode, increases due to increasing bond angle. In ( c ) the stretching frequency of H in Si–H is shifted to 2090 cm$^{-1}$ . In ( d ) the Si–OH frequency can also shift due presence of oxygen to the same silicon atom. In figure 6. ( e ) the possible influence of O is demonstrated, it indicates if number of oxygen atom bonded to any Si atom increases then dihedral angle at O will change.

## 4.1. Oxygen-Incorporation

Oxygen incorporation into a–Si network is controlled by flow variation of carbon dioxide into the deposition chamber as the shown in figure 2. Oxygen incorporation in a–Si–network may be increased by lowering chamber pressure during deposition, increasing the RF power dissipation through the source gases, lowering the $SiH_4$ + $CO_2$ flow rate or controlling the substrate temperature.

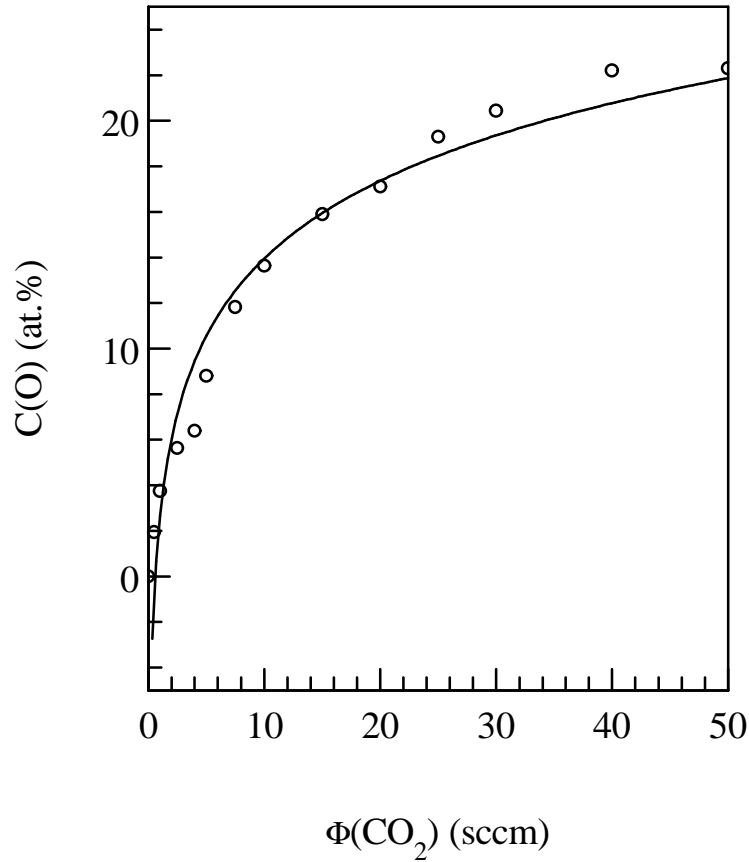

**Figure 2.** Variation of Oxygen content, in atomic percent ( C( O ) at. % ), due to $CO_2$ flow variation during deposition, for $SiH_4$ = 10 sccm.Total flow rates of $SiH_4$, $CO_2$, $H_2$ is kept at 100 sccm. The solid line is the best fit curve that can be drawn with the data points. The solid line is drawn to indicate most probable value of C( O ) at different $CO_2$ flow rates. As flow rate does not have precise influence on plasma kinetics C( O ) varies quite randomly from most expected value. Even then, oxygen content of the film is dependent upon carbon–dioxide flow.

Here total flow rate of $SiH_4$ + $CO_2$ + $H_2$ is kept at 100 standard cm$^3$ per minute ( sccm ) and $SiH_4$ at constant 10 sccm but the $CO_2$ and $H_2$ flow rates are regulated. With increasing $CO_2$ flow, r = $CO_2$/$SiH_4$ flow ratio will rise. As we go on increasing $CO_2$ flow from zero the rate of O incorporation into the a–Si network is observed to be rapid and linear (approximately and on the average, 1.2 at.% rise in C( O ) per sccm $CO_2$ flow increase ) at the initial stage, for $CO_2$ flow lower than 10 sccm. After that C( O ) does not rise at this rate. As the figure 2 shows the rate of change of C( O ) falls for $CO_2$ > 20 sccm.

Initially when carbon dioxide flow is increased but kept lower than 10 sccm and hydrogen flow higher than 80 sccm the dissociation of $CO_2$ is dominated by forward process. $H_2$ breaks into H + H in the plasma, which helps breaking $CO_2$ in CO + OH + H. OH concentration in gas phase rises. While it comes in contact with film growing surface forms Si–OH. This is the first stage of most probable mechanism of oxygen incorporation within a–Si network. H of this Si–OH may be taken away by another incoming H forming Si–O– + $H_2$ and make Si–O– site exposed to incoming $SiH_3$ radicals to form Si–O–$SiH_3$.

## 4.1.1. Higher Oxides of Si

The deposition process does not have any control on the selection of a certain group of atoms to get bonded preferentially in a particular arrangement. A Si–atom may have one, two or three neighbouring oxygen atoms in a–SiO:H. But two or three O–neighbours of any Si atom will induce increased asymmetric stretching vibration frequency, widen angle between the bonds of the concerned oxygen atom/s.

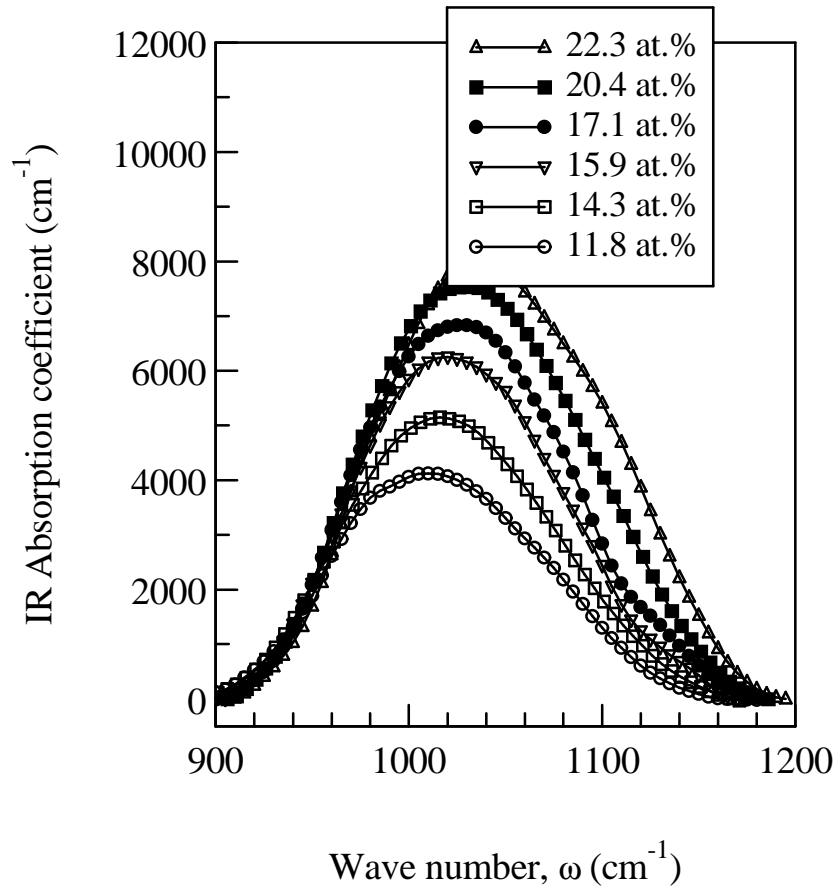

**Figure 3.** IR absorption curves within 900 cm$^{-1}$ to 1200 cm$^{-1}$ frequency region.This absorption is due to asymmetric stretching vibration of oxygen in Si–O–Si. It is observed, on increase in oxygen content absorption coefficient increases and spectra becomes broader towards higher wave number, implying that more and more higher oxides are formed, with increase in C( O ).

The rise in asymmetric stretching vibration frequency of O in Si–O–Si site gives rise to gradual appearence of hump in higher wavenumber region within 900–1200 cm$^{-1}$ band. It increases with larger incorporation of oxygen within a–Si network, (figure 3 ). The absorption peak position gradually shifts towards higher wave number with increasing peak area or absorption intensity, see figure 6. 4 also. In lower wave number region the absorption curve of figure 3, keeps its shape almost unchanged. As peak hight increases it accompanies with peak shift and becomes more flat towards higher wave number.

C( O ) is defined as the atomic percent of oxygen content within a–Si network. Higher C( O ) implies average number of oxygen atom bonded to silicon atom becomes higher for each hundred Si–atoms. In other words, average number of O bonded to one Si atom also increases. So at higher C( O ) it is probable that more Si atoms become bonded to more than one O–atom due to statistical deviation. If two oxygen atoms be bonded to a Si atom, it contributes towards 1020 cm$^{-1}$ absorption peak, for three atoms the contribution is towards 1060 cm$^{-1}$ and for four it is close to 1120 cm$^{-1}$.

These are the constituent absorption band peak positions, may be obtained by deconvoluting for constituent peaks.

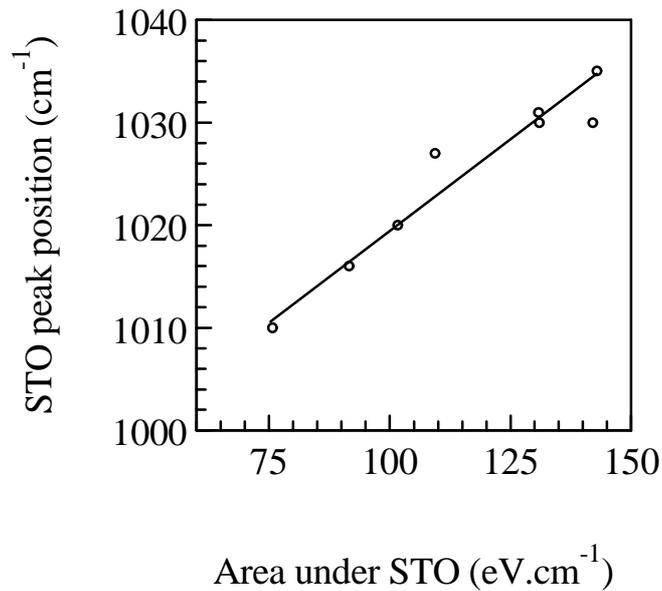

**Figure 4.** As oxygen content of the film rises the STO peak position shifts to higher wavenumber. The STO is an overall effect of asymmetric stretching vibration of oxygen in different atomic environment. Statistically it is possible that with increasing oxygen ( in atomic percent ) higher oxides of silicon will be formed. These higher oxides leads shifting the absorption peak position.

So this broadening in absorption curves at higher wave number side is due to rise in number of Si', where Si'–Si$_{(4-n)}$O$_n$: for n = 2, 3, 4.

## 4.2. Oxygen Bending Mode (BO)

It is observed that oxygen is most commonly bonded to two nearby Si–atoms forming a Si–O–Si type structure ( Mozzi, et. al. 1969 ). The array of atoms in Si–O–Si is not linear but the bond angle of O may vary from 120$^{\circ}$ to 180$^{\circ}$ ( Mozzi et. al. 1969 ). Due to non–linear shape of the Si–O–Si site and varying angle $\theta$ the oxygen atom exhibits different bond bending oscillation. These BO absorption band is shown in figure 5, for films having different oxygen content.

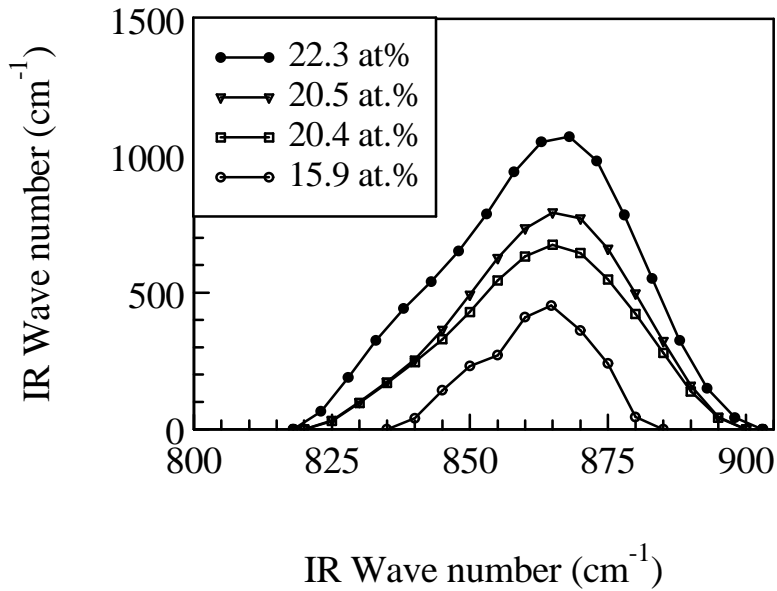

**Figure 5.** IR absorption curves of a–SiO:H films within 815 cm$^{-1}$ and 905 cm$^{-1}$ region ( BO ). The different curves are for films having different oxygen content, C( O ). As the peak height increases the curves become broader, implying the possible angular variations of bond formation at O in Si–O–Si, as C( O ) rises.

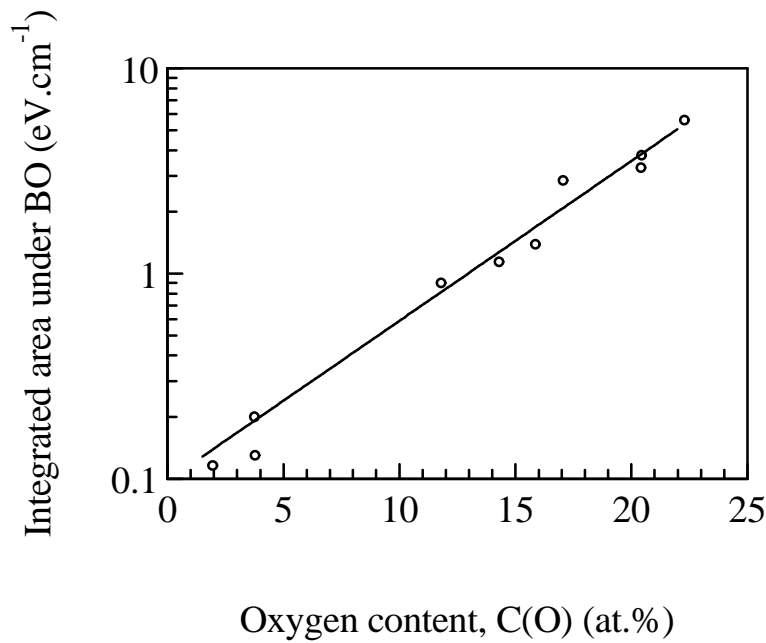

**Figure 6.** Integrated ( IR absorption band ) area, due to in–plane bending mode vibration, ( BO ), of oxygen atom at Si–O–Si site, is observed to be increasing exponentially with total oxygen content C( O ) of the film.

The nature of variation of BO with C( H ) ( not shown in figure ) is not systematic, indicating no definite direct relationship between the H atoms bonded to Si to this band of IR absorption. But as $CO_2$ flow is increased, BO is rising exponentially to C( O ), figure 6. In figure 7 it is shown how absorption constant in BO mode is related to that of STO one. The relative flat region of figure 7, for C( O ) upto 11.8 at.% may be due to bond angle $\theta$ variation. Upto this C( O ), $\theta$ may be close to 120º and beyond that it is near about 180º, figure 1.

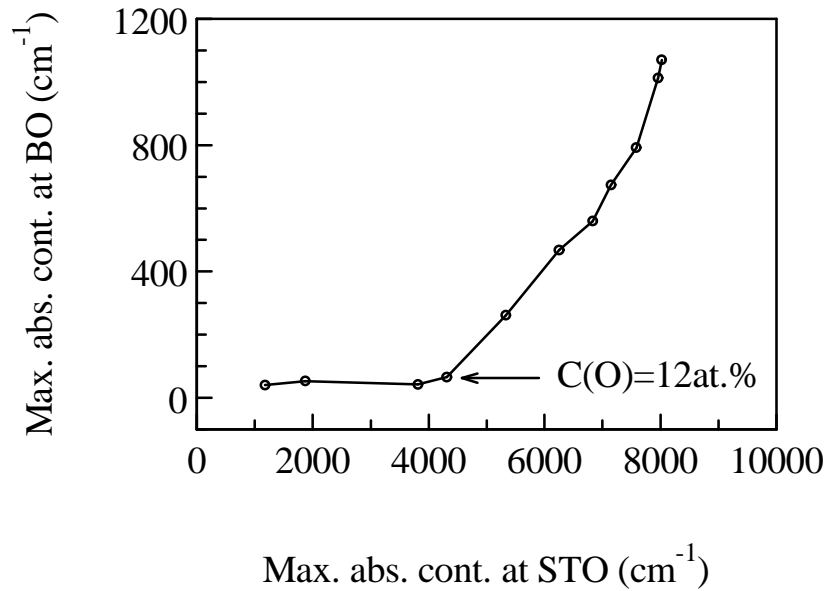

**Figure 7.** Absorption constant due to bending oscillation of oxygen is found related to that of its asymmetric stretching mode. Two regions are visible in the plot. One is when C( O ) is below 11.8 at.% or, maximum absorption constant at STO lower than 4500 cm$^{-1}$ , other is beyond it. At lower C( O ) the slower variation of BO absorption constant is due to θ ~ 120º and its faster rise if BO is due to θ ~ 180º. In these two cases the amplitude of bending vibration of O will be lower and higher respectively.

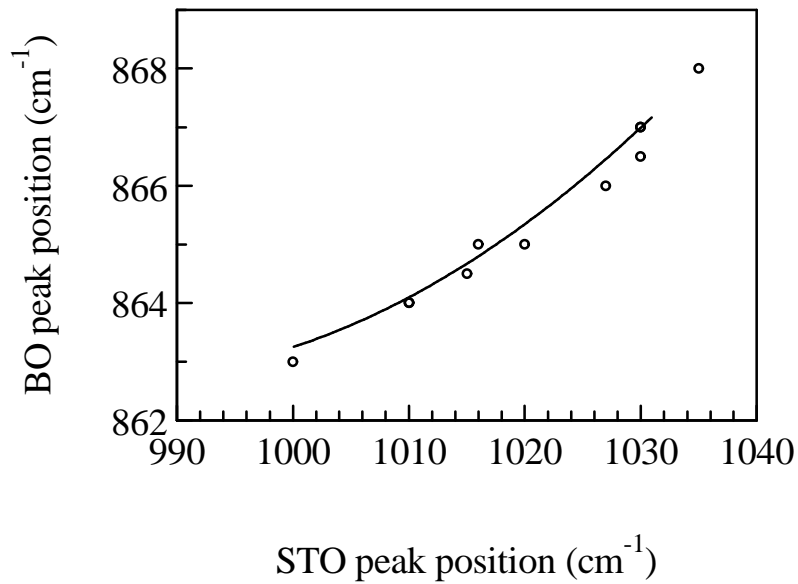

**Figure 8.** This plot shows how peak absorption position due to stretching vibration of O and its bending one depends on each other. The dotted line is drawn through, to indicate most probable variation of BO with that of STO.

Absorption peak position is an indication of local atomic arrangements or modified bonding force constant . The relationship drawn by dotted line implies STO and BO absorption are due to same atomic site, which is explained further in the text.

At higher oxygen content the bond angle at the `bridging' O rises because of possibility of more oxygen atom bonded to same Si site. Due to higher electronegativity of O, as compared to that of Si, and two unbonded p–orbitals of oxygen, the O of –O–Si–O– gets its bond angle θ to increase.

Amplitude of bending vibration of O in the plane containing Si–O–Si increases due to this increased bond angle θ . At higher C( O ) not only vibrating O–site increases in number but also higher θ causes its amplitude of vibration to enhance and resulting 860 cm$^{-1}$ absorption band to rise non–linearly with C( O ).

With changing oxygen environment, STO–peak shifts, see figure 3, which is largely due to modification of stretching force constant in Si–O bond. This modification comes up due to presence of other oxygen atom to the same silicon site. Similiarly there will be a rise in Si–O bending force constant, which will result in BO peak to shift. The peak positions of figure 5 is plotted against that of figure 3. and is shown in figure 8.

It implies BO bond is due to presence of Si–O–Si bond and also that stretching force constant modification is about 50 times more than that of bending one.

According to Tsu et. al. 1989, this might be due to Si–H bending vibration and shift of peak from 790 cm$^{-1}$ to 875 cm$^{-1}$. As shown in figure 8 the BO absorption band peak position do not shift by more than 6 cm$^{-1}$ although C( O ) increases from zero to 22 at.%. That much large shift is not visible.

So C( O ) is responsible for the 865 cm$^{-1}$ IR absorption band intensity, fig. 6. This intensity depends upon number of vibrating O atoms and also amplitude of vibration of each O. The figure 5 shows that as C( O ) increases the 865 cm$^{-1}$ absorption increases along with peak broadening. At higher bonded oxygen to the amorphous network the randomness of bond formation at O is increased due to its dihedral angle variation from 120º to 180º ( Mozzi et. al. 1969 ) in a random manner. Figure 6. 5 shows such randomness increases due to higher O–incorporation.

## 4.3. Si-OH formation

It is found difficult to isolate 3550 cm$^{-1}$ absorption peak as due to hydrogen stretching mode in SiO–H from H–O–H or water molecule ( that comes from adsorbed water vapour within the film ), as the latter has resonance vibrational frequency due to oscillation of H very close to 3550 cm$^{-1}$.

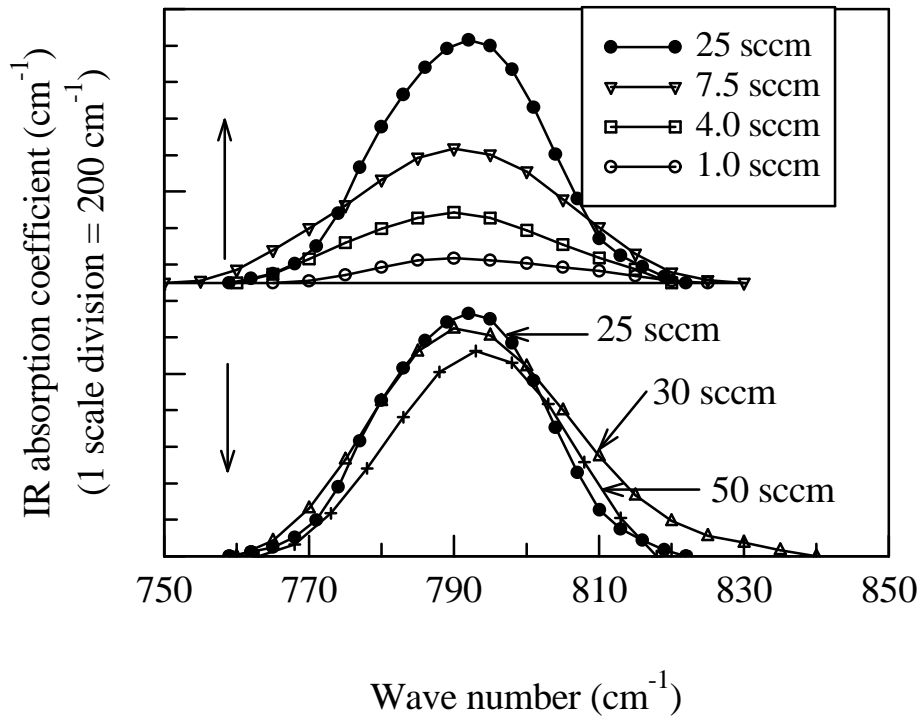

**Figure 9.** IR absorption curves due to stretching mode of oscillation of –OH in Si–OH. This absorption band lies within 750 cm$^{-1}$ and 840 cm$^{-1}$. Different curves film that are prepared with different carbon dioxide flow rates.

OH of Si–OH oscillates with a combined mass of O and H, due to Si–O stretching force constant of 6.043×10[5] dyn/cm ( Venkateswarlu, et. al. 1955 ) gives rise to absorption peak close to 776 cm$^{-1}$ which in our sample is observed around 778 cm$^{-1}$, figure 9.

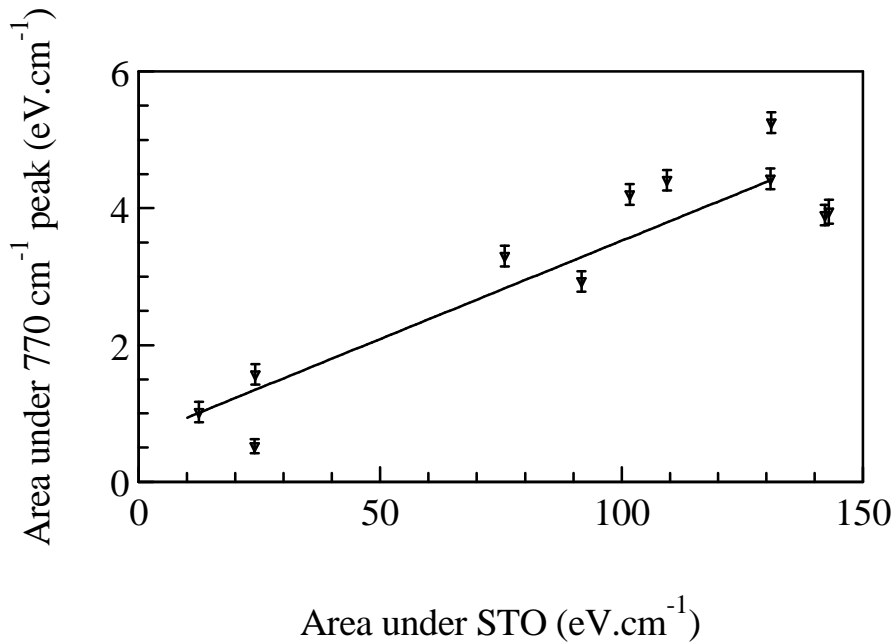

**Figure 10.** Area under 778 cm$^{-1}$ ( due to Si–OH ) absorption band is plotted to that against STO. Si–OH formation preceeds to Si–O–Si formation. Part of Si–OH gets converted into Si–O–Si and rest remains unchanged. The average straight line that is drawn, indicates there is, on the average, a certain fraction of total Si–OH that is formed, gets converted into Si–O–Si.

There is a high possibility that this band ~778 cm$^{-1}$ arises due to Si–H bending mode of oscillation ( Tsu, et. al. 1989 ) influenced by oxygen. In figure 10 it is observed that this absorption strength and that within 900–1200 cm$^{-1}$ are related linearly or otherway. The straight line is drawn to indicate the possible average variation or relationship due to increased Si–O–Si formation. The fluctuation from average tendency is mostly due to plasma processes. For a particular deposition condition one precursor may be active compared to the other. In another condition another plasma process may be dominant. But as shown in the figure 6. 10 the average tendency is as described previously. The type of variation is in accordance with the depositon mechanism stated in ( A ). Si–OH formation is the first step towards attaining Si–O–Si structure. Statistically part of this Si–OH remains unchanged in structure. Conversion of Si–OH into Si–O– hence Si–O–Si is again dependent upon H atoms available near film growing surface.

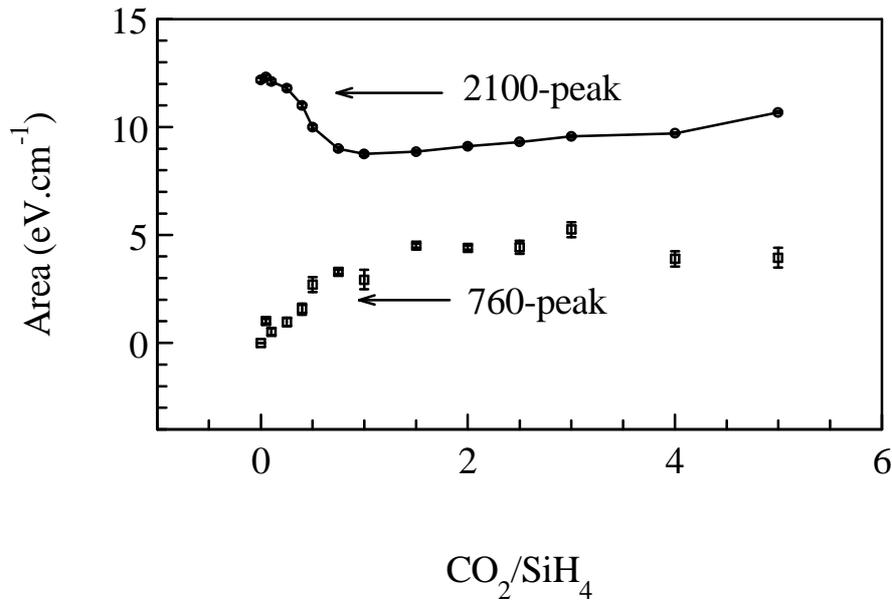

**Figure 11.** The curves are drawn to show how hydrogen incorporation within a–SiO:H films varies in Si–OH form as well as Si–H$_n$ ( where n = 1,2 or 3) due to CO$_2$/SiH$_4$ flow ratio variation. The `area' is the integrated area of the absorption coefficient under wave number axis. Curve ( i ) is the integrated area of 760 cm$^{-1}$ IR absorption band due to oscillation of OH in Si–OH, denoted by `760–area'. Curve ( ii ) is the integrated area of 1850 cm$^{-1}$ to 2250 cm$^{-1}$ IR absorption band that is due to oscillation of H in SiH$_n$ indicated as `2100–area', in the figure.

The integrated absorption strength when plotted against CO$_2$ flow rate shows steady rise from 0 to 4.5 eV.cm$^{-1}$ untill CO$_2$ = 15 sccm, figure 6. 11 curve ( i ). At this point H$_2$ flow is higher than 70 sccm but lower than 90 sccm.

## 4.4. Modification in Si-H vibration

As CO$_2$ flow rate is increased, integrated absorption area under 1850 to 2250 cm$^{-1}$ band is observed to fall steadily for CO$_2$ flow upto 10 sccm, figure 11 curve ( ii ). Beyond this point no such rapid change is observed. That is initially C( H ) falls with rise in CO$_2$ flow. Hydrogen incorporation in a–Si network depends upon [H]. Due to the presence of CO$_2$ during deposition part of this H is lost in the form of OH. At increased CO$_2$ and lowered H$_2$ flow in RF plasma region, [OH]/[H] increases. So more Si–OH is formed, figure 11 curve ( i ), partly in expense of Si–H, figure 11 curve ( ii ).

Incorporated O into the a–Si network induces the Si–H stretching frequency to shift from 1990 cm$^{-1}$ to 2090 cm$^{-1}$, ( see figure 12 ). Absorption value at 1990 cm$^{-1}$ diminishes and that at 2090 cm$^{-1}$ rises up. This 2090 cm$^{-1}$ IR absorption is primerily due to stretching vibration of Si–H influenced by oxygen atom bonded to this Si in Si–O–Si form. Position of maximum absorption in STO mode is an indication of most probable oxygen environment of Si. When this oxygen environment puts inductive effect on Si–H stretching force constant then its absorption constant is bound to be reflected at the 2090 cm$^{-1}$ IR responce. As demonstrated in figure 6. 12, the 2090 cm$^{-1}$ absorption depends upon oxygen content of the film. Whereas figure 11 curve ( ii ) indicates total hydrogen content of the film falls at higher CO$_2$ flow. So in the presence of oxygen within a–Si network, evolution of 2090 cm$^{-1}$ peak is due to Si–H stretching vibration influenced by inductive effect of oxygen bonded to the same Si atom.

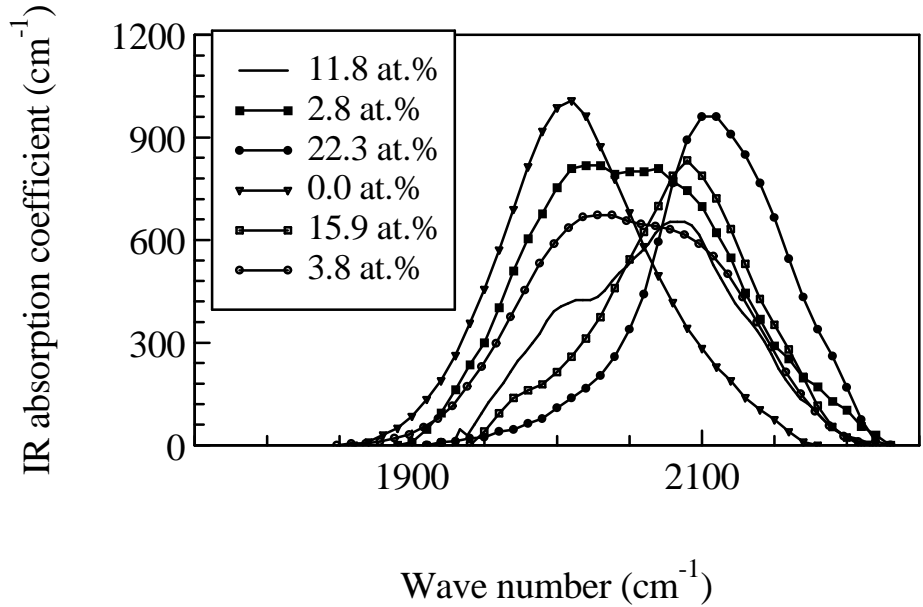

**Figure 12.** This figure shows the IR absorption curves of the a–SiO:H films, the absorption coefficient of which is shown within the region of 1850 cm$^{-1}$ to 2250 cm$^{-1}$. Different curves are drawn superimposing one on another. As oxygen content within the film increases 1990 cm$^{-1}$ peak diminishes, whereas 2090 cm$^{-1}$ one rises.

This absorption peak at 2090 cm$^{-1}$ is not due to SiH$_2$ structures but the Si–H centres where the Si– has at least one O atom back bonded to it ( Hubner 1980 ).

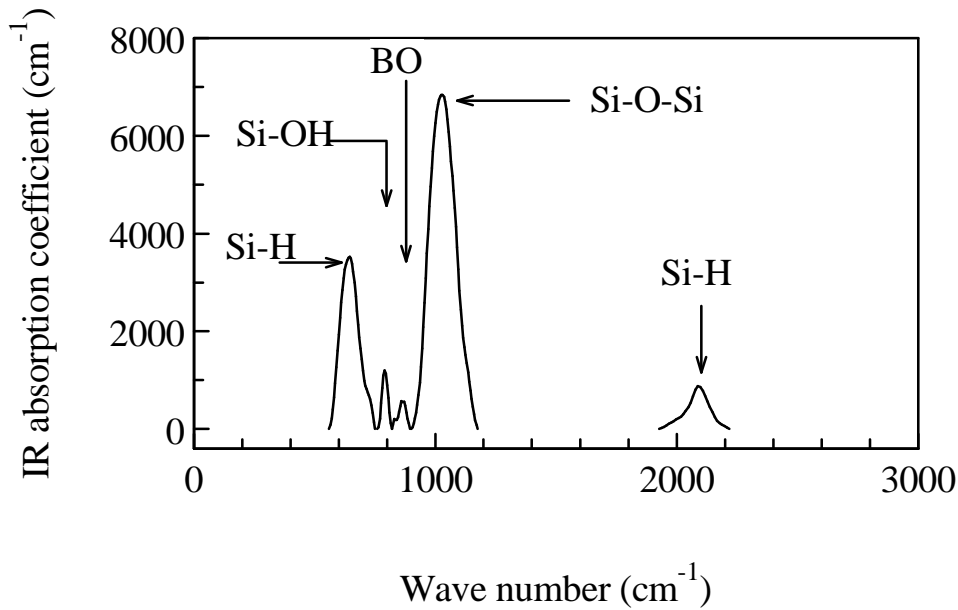

**Figure 13.** Full range IR absorption spectra of a–SiO:H films deposited on crystalline Si substrate. The spectra is taken within 400 to 4400 cm$^{-1}$ region with an instrumental resolution of 1 cm$^{-1}$.

Initially when SiH$_4$ = 10 sccm, CO$_2$ = 0.0 sccm H$_2$ = 90 sccm the possibility of oxygen incorporation in a–Si network is very low and the Si–H stretching absorption band shows little or no $SiH_2$ related absorption peak around 2090 cm$^{-1}$. When $H_2$ flow rate is further decreased the $SiH_2$ formation probability becomes still lower. Part of H atoms, while reacting with $CO_2$ , are transformed into OH, and instead of extra H, OH gets bonded to Si–atoms at the film growing surface.

Hydrogen dilution during deposition is used to minimize unfulfilled bonds of Si. During a–SiO:H film preparation these dangling bonds of Si are filled up increasingly by OH when $CO_2$ is increased, and less H is bonded to Si atoms. Figure 9 shows 750 – 840 cm$^{-1}$ IR absorption band rises with $CO_2$ flow, that is, more Si–OH bonds are formed than that of Si–H or $SiH_2$. So $SiH_2$ formation probability within the film lowers with rise in $CO_2$ flow.

While oxygen incorporation is increased H atoms in monohydride configuration are prefarentially attached to those Si atoms that has Si–O bond.

One full range spectra is shown in figure 13, where IR absorption is measured from 400 to 4400 cm$^{-1}$ region.

## § 5. Discussions

Incorporated oxygen in a–SiO:H films is increased by increasing $CO_2$ flow or at higher r. At higher r more $CO_2$ molecules break into CO + OH, which eventually leads to incorporated oxygen to rise.

$SiH_4$ breaks in RF plasma by $SiH_3$ or/and $SiH_2$ radicals where $SiH_3$ is a desirable precursor for good quality film preparation. Whereas breaking of $CO_2$ is dependent on hydrogen atom density within the plasma.

Atomic percent of incorporated oxygen is measured from 900 to 1200 cm$^{-1}$ IR absorption spectra. In this region the absorption takes place due to Si–O–Si type structure. Whereas, as shown previously, incorporated oxygen atom may be present in Si–OH form also. Carbon monoxide that is formed in RF plasma may partly be incorporated within the film but IR absorption band assignment of this incorporated CO is difficult.

CO may be bonded to any one Si atom by forming a double bond with it or it may be linked with two separate Si atoms. In the two cases C = O stretching force constant is different. In the first case reduced mass of oscillation will be different from individual mass of O or C atom. Carbon atom is bonded to the a–Si network at the other end of the double bond, so it is not totally free to oscillate and what will be the exact reduced mass, is difficult to state. Hence it is highly unpredictable in which region the frequency of vibration will lie. Similiarly in the latter situation the C atom is more firmly held within the network, and its frequency of vibration although may be assigned but it will not give total CO content of the film.

So C( O ) gives the atomic percent of oxygen bonded in Si–O–Si form, it does not indicate total oxygen content of the film.

Similiarly C( H ) does not mean total hydrogen content of the film. A large part of hydrogen is found bonded in Si–OH form.

Hydrogen atoms are bonded to Si and oxygen atoms quite randomly but in a definite atomic proportion. At higher oxygen availability within the network, hydrogen content in Si–H form falls but Si–OH one rises.

If we see the electronic properties of a–SiO:H films, Si–H or Si–OH bonds are not very much deteriorative, instead they may be called as free bond–terminators. But Si–O–Si structure is chain like within the amorphous network. Whenever there is an electronic wavefunction travelling from one Si atom to another, this Si–O–Si will give rise to an obstacle by separating the two Si atoms further apart and also creating slightly electron rich centre around O. If the angle between the two bonds of O, $\theta$ , be more, the Si atoms will be even further apart and the result is yet worse electronic conduction phenomena.

This is the reason as area under BO increases, electronic property deteriorates further. The angle $\theta$ between the bonds of O is not definite. It cannot be found from IR spectra, but it indicates that $\theta$ does vary and its effect on material property is dominating. This disordered bond formation of oxygen will have a major influence on its bending vibration. The IR absorption band of BO indicates this disorder increases at higher concentration of bonded oxygen.

## § 6. Relation between structural and optoelectronic properties

Selection of a photosensitive amorphous semiconductor in the active layer of amorphous silicon solar cells can approximately be scaled from the following parameters; activation energy ( $\Delta E$ ) and darkconductivity prefactor ( $\sigma_0$ ), in the expression ( 1 ), Tauc's gap ( Eg ) in the formula ( 2 ) and photoconductivity $\sigma_{ph}$.

These optoelectronic parameters gives an indication of how effective the material is as a photosensitive alloy–semiconductor.

### 6.1. Effect of C(O)

The major change that takes place during the above mentioned sample preparation is incorporation of oxygen, C( O ), within a–Si network. It is deteriorative in the sense that more oxygen in Si–O–Si configuration reduces electronic conduction, e.g., secondary photoconductivity, $\sigma_{ph}$. In Table 1, the effect is demonstrated. Lower $\sigma_{ph}$ is an undesirable property of the material for its use in photovoltaic device. But for multijunction solar cell high optical gap of intrinsic layer is one of the major criteria for its use in top cell. As indicated in the table, Eg increases with oxygen content. Among the two opposing tendencies one optimum condition may be chosen, depending upon device requirements.

$\sigma_0$ is equivalent to minimum metallic conductivity, $\sigma_{min}$ , and its reduction indicates interaction length ( `a' ) through which electronic wave functions interact ( Mott and Davis 1979 ). At higher C( O ) or lower $\sigma_O$ values, the material will have higher value of `a'. If Si sites be the atomic positions through which conduction takes place then according to figure 1, it increases with increase in $\theta$ at higher C( O ).

### 6.2. Effect of BO

The 866 cm$^{-1}$ IR absorption band which reveals the characteristic in plane bending mode of oscillation of oxygen in Si–O–Si, has a dominant influence on optoelectronic properties. It may vary for a particular level of oxygen content within the film. This is not contrary to what is shown in figure 6, because this figure is for a particular set of samples having same rf power density, deposition pressure, substrate temperature. As it is observed the magnitude of BO varies along with C( O ). So we had to choose from other set of samples to compare their properties for a particular C( O ) but with different intensity of BO.

**Table 1:** Dependence of optoelectronic properties of a–SiO:H films on atomic oxygen content in Si–O–Si form, C( O ). Deposition temperature of these films are $200^o$ C. With increase in C( O ) the value Eg, $\sigma_{ph}$, $\Delta E$, $\sigma_0$ falls.

| Sample | C( O )<br>(at.%) | Eg<br>(eV) | $\sigma_{ph}$<br>(S.cm$^{-1}$) | $\Delta E$<br>(eV) | $\sigma_0$<br>(S.cm$^{-1}$) |
|--------|--------|--------|--------|--------|--------|
| 30 | 8.92 | 1.96 | $1.35 \times 10^{-6}$ | 1.01 | $5.2 \times 10^{5}$ |
| 29 | 13.77 | 1.96 | $1.95 \times 10^{-6}$ | 0.96 | $1.39 \times 10^{5}$ |
| 18 | 17.03 | 2.00 | $3.67 \times 10^{-7}$ | 1.00 | $1.25 \times 10^{5}$ |
| 20 | 19.81 | 2.06 | $4.89 \times 10^{-8}$ | 0.97 | $1.57 \times 10^{4}$ |
| 34 | 21.97 | 2.05 | $1.26 \times 10^{-8}$ | 0.89 | $3.92 \times 10^{2}$ |

In Tables 2 and 3 it is observed that as BO increases at a constant C( O ), the film quality deteriorates by lowering of photoconductivity, dark conductivity activation energy and $\sigma_0$. At higher BO corresponding Eg is observed to increase even at constant C( O ) but $\sigma_{ph}$ falls, lowering of $\sigma_0$ is also observed. In Table 1, when C( O ) increases then according to figure 6, BO will also rise. This Table is arranged in decreasing order of $\sigma_{ph}$. So it is visible that increase in BO is deteriorative to film properties.

**Table 2:** Dependence of optoelectronic properties of a–SiO:H films on IR detected BO mode of absorption ~865 cm$^{-1}$. Deposition temperature of these films is $200^o$ C.

C( O ) = ( 20.05 ± 0.01 ) at.%

| Sample | BO<br>(eV/cm) | Eg<br>(eV) | $\sigma_{ph}$<br>(S.cm$^{-1}$) | $\Delta E$<br>(eV) | $\sigma_0$<br>(S.cm$^{-1}$) |
|--------|--------|--------|--------|--------|--------|
| 24 | 2.85 | 1.99 | $1.89 \times 10^{-6}$ | 0.92 | $4.2 \times 10^{4}$ |
| 16 | 3.78 | 2.06 | $1.17 \times 10^{-8}$ | 0.90 | $6.1 \times 10^{2}$ |

## 6.3. Effect of OH formation

Visibly, presence of OH within the a–SiO:H films does not show any deterioration in optoelectronic properties. Amount of OH formation depends on deposition conditions but the dominant effect of oxygen stretching vibration within Si–O–Si and of bending mode of oscillation, masks the effect of OH on optoelectronic properties.

**Table 3:** Dependence of optoelectronic properties of a–SiO:H films on IR detected BO mode of absorption ~865 cm$^{-1}$. Deposition temperature of these films is 200 $^{0}$C.
C( O ) = ( 20.15 $\pm$ 0.01 ) at.%

| Sample | BO (eV/cm) | Eg (eV) | $\sigma_{ph}$ (S.cm$^{-1}$) | $\Delta E$ (eV) | $\sigma_0$ (S.cm$^{-1}$) |
|---|---|---|---|---|---|
| 26 | 3.13 | 2.07 | $5.21 \times 10^{-8}$ | 0.88 | $1.6 \times 10^{3}$ |
| 28 | 3.84 | 2.12 | $7.17 \times 10^{-9}$ | 0.86 | $2.1 \times 10^{2}$ |

OH is a silicon dangling bond terminator just like H. So its deteriorative effect on electronic property may not be expected, which may also be the reason why distinctive effect of Si–OH on electronic property is not visible.

## § 7. Conclusions

Oxygen is a Group VI element. It has valency two. Two bonds of it can have any angle between 120º to 180º. Hence the amorphous network of hydrogenated amorphous silicon oxide, gets tetrahedral Si bonds plus dihedral oxygen bonds. Short range order like bond angle at each atomic site is not strictly followed and the material is subjected to irregular structure. The electronic wave function has to interact with one more irregularity; like short range disorder. We point out this in terms of the observable BO. With increasing BO the effective correlation length should rise and is supported by lowering in $\sigma_{min}$ or $\sigma_0$.

The a–SiO:H films deposited in rf glow discharge deposition system, reveal certain characteristic structural configurations, some of which are desirable and some are not. The structural studies done by FTIR spectrophotometry show a good correlation of structure with optical and electronic properties

( i ) Local atomic structure changes with different deposition conditions

( ii ) Si–H bond stretching force constant increases due to oxygen being back bonded to Si at higher C( O ). Stretching force constant of Si–H bond in a–Si:H is 2.50×10$^5$ dyn/cm, but here it is observed that this force constant is increased to a maximum of 2.60×10$^5$ dyn/cm, corresponding to 2090 cm$^{-1}$ peak position. H is preferentially bonded to those Si that has oxygen atom bonded to it.

( iii ) Si–OH stretching vibrational absorption band is observed close to 778 cm$^{-1}$ wavenumber. But this OH formation within the films, exhibits no definite influence on film properties.

( iv ) 866 cm$^{-1}$ absorption band is assigned to in–plane–bending mode of oscillation of oxygen atoms, it rises exponentially with rise in C( O ), and has deteriorating effect on film properties.

It is expected that if it is possible to control these defects, the a–SiO:H films thus prepared will be of superior quality.